\documentclass[journal]{IEEEtran}

\ifCLASSINFOpdf
  \usepackage[pdftex]{graphicx}
  \graphicspath{{fig/}}
  \DeclareGraphicsExtensions{.pdf,.jpeg,.png}
\else
  \usepackage[dvips]{graphicx}
  \graphicspath{{fig/}}
  \DeclareGraphicsExtensions{.eps}
\fi

\usepackage{url}     

\usepackage{algorithm}
\usepackage{algorithmicx}
\usepackage{algpseudocode}

\usepackage{color}

\begin{document}

\title{An Extended Epidemic Model on Interconnected Networks for COVID-19 to Explore the Epidemic Dynamics}

\author{OU~DENG,~\IEEEmembership{Member,~IEEE, }
        KIICHI~TAGO,~\IEEEmembership{Member,~IEEE, }
        and~QUN~JIN,~\IEEEmembership{Senior~member,~IEEE}

\thanks{Ou Deng, is with the Graduate School of Human Sciences, Waseda University, 2-579-15 Mikajima, Tokorozawa 359-1192, Japan (e-mail: dengou@toki.waseda.jp).}
\thanks{Kiichi Tago is with the Department of Information and Network Science, Chiba Institute of Technology, 2-17-1 Tsudanuma, Narashino, Chiba, Japan (e-mail: kiichi.tg@aoni.waseda.jp).}
\thanks{Qun Jin is with the Graduate School of Human Sciences, Waseda University, 2-579-15 Mikajima, Tokorozawa 359-1192, Japan (corresponding author, phone: +81-4-2947-6906; e-mail: jin@waseda.jp).}}



\maketitle

\begin{abstract}
COVID-19 has resulted in a public health global crisis. The pandemic control necessitates epidemic models that capture the trends and impacts on infectious individuals. Many exciting models can implement this but they lack practical interpretability. This study combines the epidemiological and network theories and proposes a framework with causal interpretability in response to this issue. This framework consists of an extended epidemic model in interconnected networks and a dynamic structure that has major human mobility. The networked causal analysis focuses on the stochastic processing mechanism. It highlights the social infectivity as the intervention estimator between the observable effect (the number of daily new cases) and unobservable causes (the number of infectious persons). According to an experiment on the dataset for Tokyo metropolitan areas, the computational results indicate the propagation features of the symptomatic and asymptomatic infectious persons. These new spatiotemporal findings can be beneficial for policy decision making.
\end{abstract}


%
\IEEEpeerreviewmaketitle

\section{Introduction}

\IEEEPARstart{T}{he} presence of COVID-19 has resulted in a global health crisis. Controlling the pandemic requires monitoring and evaluating the infection transition, and this is an aspect that direct statistics are unable to measure. This study presents an approach that infers the infectious population as target unobservable information from the limited observable factors. This includes the daily confirmed cases, incubation period, and spreading routes.

The SEIR model is a classic epidemic model that consists of four classes of susceptible (S), exposed (E), infected (I), and recovered (R) individuals. Here, S+E+I+R=N, where N is the total population of a region that can be either a country, a metropolitan area, or a city. In the proposed network model, N is defined as the number of nodes.

When considering the new features of COVID-19, we extended the SEIR model. First, we included the asymptomatic (A) component, which is the group of infected individuals that are asymptomatic. Consequently, we defined E, I, and A as the  {\it{Infectious Classes}} in our model. Their demographic transition is the main target of the unobservable information that is explored in this study. Second, we included the hospitalized (H) component, which consists of the new daily confirmed cases that were announced by public health authorities. Thus, the extended SEIR is referred to as the SEIRAH model.

The real world is similar to interconnected networks, and the constituents create new paths of propagation \cite{Newman2006}. Previous studies have clarified that the propagation is significantly affected by the structure of the social networks and human mobility\cite{Gama}\cite{NWK-Science}.

This study proposes interconnected networks, a spatial structure with primary human mobility. A simplified analysis is performed that consists of several residential regions and one central work region, which is a typical metropolitan model. Residential regions are individual social networks with family based activities. The work region is also an individual network, but it is organized by the daily commuting of people from the residential regions. 

\begin{figure}[!t]
\centering
\includegraphics[width=0.48\textwidth]{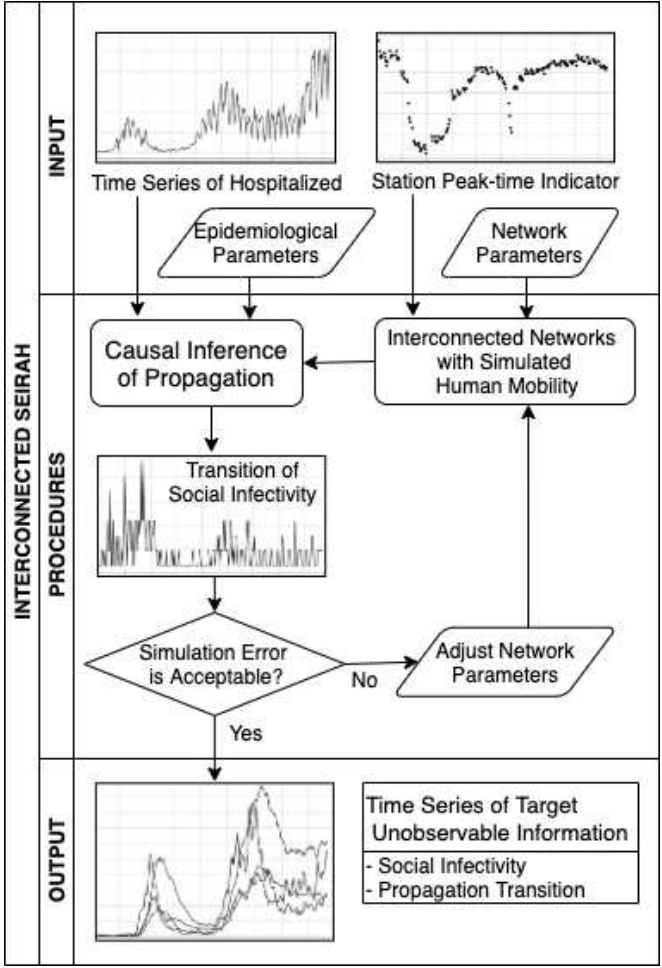}
\caption{Framework of the interconnected SEIRAH. General description of the computing procedures in this study.}
\label{fig_framework}
\end{figure}

We overlaid the SEIRAH model on the interconnected networks with the overall framework called the interconnected SEIRAH. It is the simulated society model that explores the propagation dynamically, as described in Fig.\ref{fig_framework}. In the SEIRAH model, {\it{Social Infectivity}} is defined as the probability of infection between the infectious and susceptible individuals in social activities. We can express the social infectivity as an interventions estimator of the propagation in the SEIRAH mechanism.

We built causal inferences for the social infectivity. The number of daily new cases, namely the hospitalized ($H_t$) time series, is the observable source information as the {\it{Effect of Propagation}} in the mechanism. The infectious classes consist of unobservable target information as the {\it{Causes of Propagation}}. A computational path is from the effect to the causes of the SEIRAH mechanism in the interconnected dynamical networks.

Following Section III described the structure and mechanism of the interconnected SEIRAH framework. Moreover, Section IV verified this framework using the experiential data that belongs to the Tokyo metropolitan areas since the COVID-19 outbreak. The simulation output the inference results with the findings of the epidemic dynamics.

Furthermore, a networked database can be generated by using the proposed framework's computing processes. The network nodes store the time series of the status transitions that are organized by the spatial topology. This networked database can be a value-added outcome that collaborates with classic epidemiological statistics by supporting either the macro- or micro-level observations of the propagation, even at the individual level. This methodological collaboration will help us understand more propagation features and provide more useful references to control a pandemic.

\begin{figure*}[!t]
\centering
\includegraphics[width=1.0\textwidth]{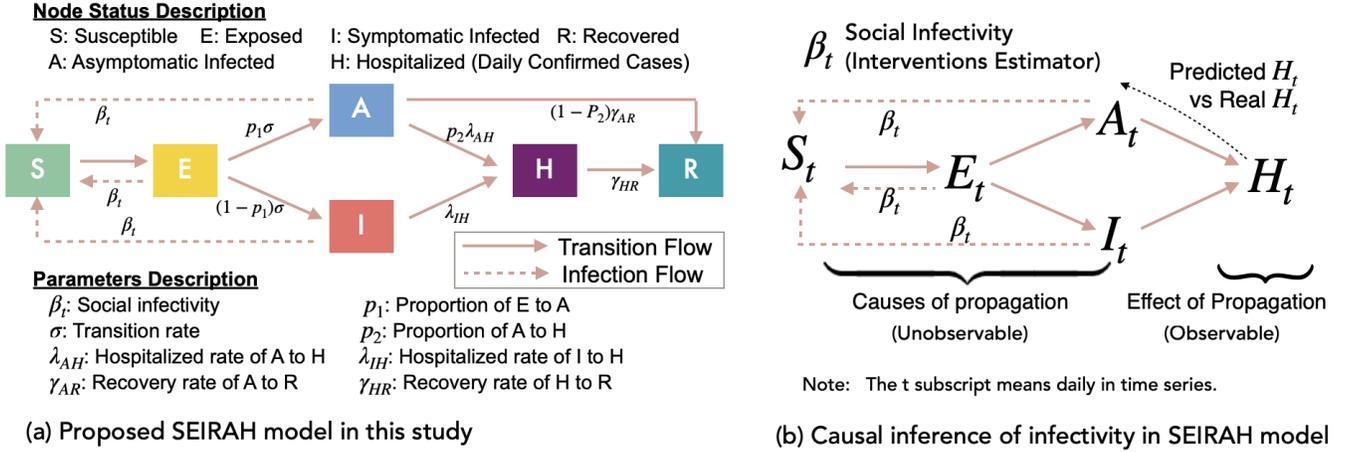}
\caption{Epidemic SEIRAH model and the propagation mechanism by the stochastic processes.}
\label{fig_SEIRAH}
\end{figure*}

\section{Related Works}

\subsection{Spatial Structure in Epidemic Dynamics}

The small-world network model, which was introduced by Watts and Strogatz in 1998 \cite{WS1998}, indicated two important properties that are observed in many societies: a small diameter and a high clustering coefficient. This milestone study tested a simplified epidemic model and determined that the propagation in small-world networks might be faster and broader than in regular networks. The small-world model has emerged in terms of network research in social sciences. Newman et al. \cite{Newman2006} developed fundamental percolation and mean-ﬁeld methods to study the propagation in small-world networks.

The severe acute respiratory syndrome (SARS) outbreak that occurred in 2003 provided a chance to verify some of the theoretical issues in the small-world. Michael and Chi \cite{SW-SARS} determined that the spreading of SARS in Hong Kong had typical small-world properties. Consequently, epidemics have focused more on the spatial factors. Gama and Nunes \cite{Gama} attempted the SEIR model in the small world in 2006. They indicated that an epidemic model might consider that the populations are ﬁnite and discrete, and they include a realistic representation of the spatial degrees of freedom or the interaction network structure.

Distance is one of the basic concepts in spatial and networked analyses. Brokmann et al. \cite{HiddenGeometry} proposed the effective distance concept of social networks in 2013. In their study, the intricate spatiotemporal patterns could be reduced to simple, homogeneous wave propagation patterns if a stochastically motivated effective distance replaces the conventional geographic distance. In the context of global, air-trafﬁc-mediated epidemics, it has been shown that an effective distance reliably predicts the disease arrival times, even if the epidemiological parameters are unknown. This approach could also identify the spatial origin of the spread of SARS in 2003, and it was successfully applied to the data of the worldwide H1N1 influenza in 2009. This indicates that other network-driven dynamic processes might also show an effective distance, such as rumors, opinions, innovations, and other social phenomena. The concept of social infectivity in our study is based on the concept of effective distance.

Recent studies have explored the interactions between the networks. Liu et al. \cite{PLos} studied epidemics in two interconnected small-world networks by using the SIS (susceptible-infected-susceptible) model. They described that the epidemic threshold in these networks decreases when the rewiring probability of the component small-world networks increases. When the infection rate is low, the rewiring probability affects the global steady-state infection density. However, when the infection rate is high, the infection density is insensitive to the rewiring probability. Moreover, they indicated that the epidemics in the interconnected small-world networks can spread at different speeds, which depends on the rewiring probability. Following their leads, this study explores the various network patterns in epidemic dynamics by adjusting the rewiring probability in the following experiment.

\subsection{Human Mobility Affects the Propagation}

Barabasi \cite{NWK-Science} pointed out that, in contrast to the pathological parameters, the social network structure and people mobility affect the propagation more signiﬁcantly.

Transportation networks and their functions during an epidemic have been recently studied. Qian et al. \cite{Transit} studied how transportation changed the spread of the epidemic in three major metropolitan areas in China. Their model explained how the structural properties of the metropolitan contact networks are associated with the risk level of communicable diseases. Their results highlighted the vulnerability of the urban mass transit systems during disease outbreaks, and they suggested important planning and operation strategies to mitigate the risk of communicable diseases.

Mobility was determined to be an essential factor in spreading epidemics. Chang et al. \cite{Mobility} investigated mobility tracks by using the smartphone location data of nearly 98 million people in ten large cities in the United States. The mobility is between the census block groups (CBGs, living areas) and the points of interest (POIs, e.g., cafe shops and restaurants), which includes the information on the staying time and the spatial size of the locations. The CBGs-POIs model simulates an interconnected structure that explores the propagation. According to the information on the daily new cases, we can conﬁrm the infection rate in different cities. Like the CBGs-POIs structure, this study developed an interconnected residence-work network model that resembles Japan's urban life pattern.

\subsection{Position of this Study}

This study presents a novel networked framework for epidemic dynamics, which includes computational methods for epidemiological trends and status transitions. The core function is the causal inference that clarifies the unobservable causes and the intervention estimator from the limited observable effects of the propagation. This extended epidemic model in interconnected networks is a value-added model that collaborates with classic epidemiological statistics that lacks spatial analysis. This methodological collaboration provides another type of networked causal analysis path to explore the epidemic dynamics.

\section{General Framework}

\subsection{SEIRAH Model and Social Infectivity}

In classic epidemic models, infectivity is one of the static epidemiological parameters from clinical statistics compatible with most local or short-term infectious diseases. Unlike these diseases, the COVID-19 pandemic has signiﬁcant social features, including a droplet infection route and an impact on human mobility. Therefore, the infectivity of COVID-19 cannot be considered to be a static parameter; instead, it is a variable parameter determined by different social circumstances.

This study extended the classic SEIR to the SEIRAH epidemic model as shown in Fig.\ref{fig_SEIRAH}(a). The three E, A, and I classes are infectious classes that are expressed as propagation causes. The H class is the effect of the propagation. The social infectivity is the interventions estimator between the causes and effects, as shown in Fig.\ref{fig_SEIRAH}(b).

\begin{figure}[!t]
\centering
\includegraphics[width=0.45\textwidth]{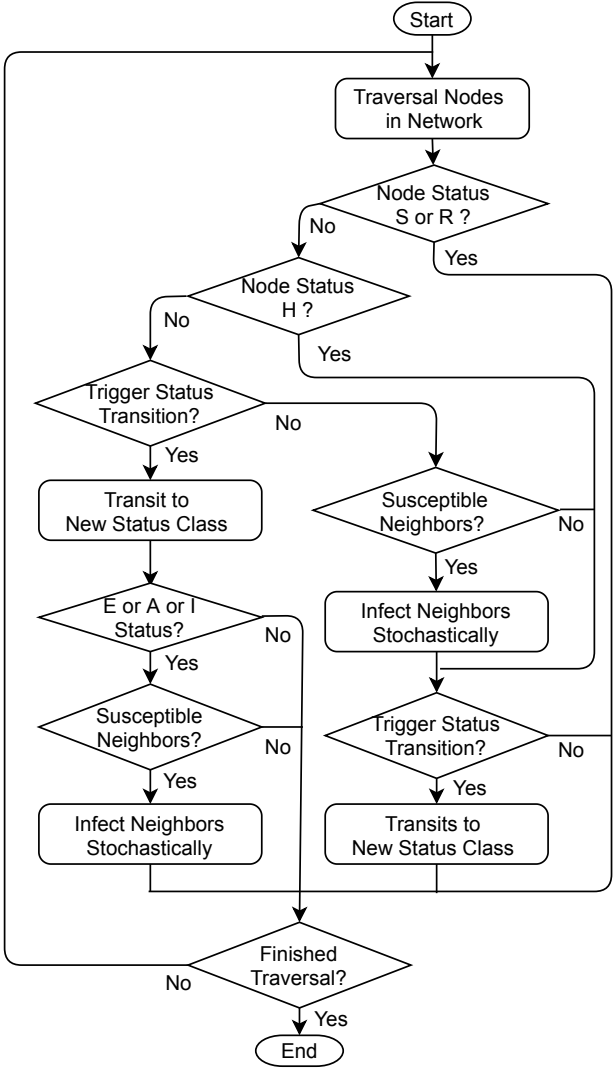}
\caption{Algorithm 2: Propagation mechanism by the stochastic processes.}
\label{fig_algorithm_2}
\end{figure}

The SEIRAH propagation mechanism is illustrated in Fig.\ref{fig_algorithm_2}. The status transition can be considered to be a stochastic process \cite{Kucharski}\cite{Threshold-1}. Therefore, in the simulation, the transition occurs when the random triggering probability exceeds a specific threshold value.

\subsection{Interconnected Networks}

\begin{figure}[!t]
\centering
\includegraphics[width=0.48\textwidth]{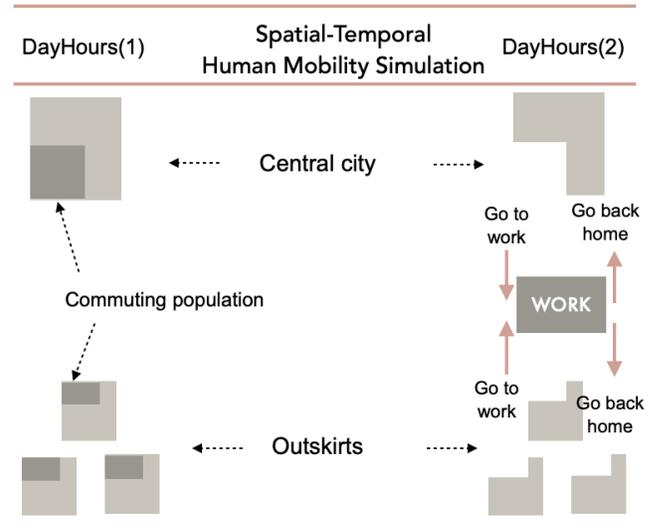}
\caption{Interconnected networks framework with major human mobility (daily commuting) in a typical metropolitan model.}
\label{fig_R+W}
\end{figure}

This study modeled society as a combination of residence-work networks in a typical metropolitan area with one center and several outskirts, and they were initialized as individual networks. The work network consists of people that are commuting daily from the residential regions. 

Here, daily commuting is considered to be patterned human mobility in the metropolitan model. The work network interconnects all the residence networks and it has no transmission among the residence networks.

This model divides one day into two time zones of `DayHours(1)' and `DayHours(2)' for a simpliﬁed daily human activity pattern. The former is the lifetime, and the latter is the working hours. These are temporal factors for the time-series analysis in the interconnected SEIRAH.

The above decomposes the human mobility simulation to a particular spatial-temporal pattern, as shown in Fig.\ref{fig_R+W}.

Then, we choose a suitable network model to simulate the real world. Small-world networks possess short node-to-node distances and a high degree of clustering, which is similar to the human community. Many previous studies indicate that the small-world effect is one of the most common consequences of disease propagation \cite{Newman2006}. We chose the Newman-Watts small-world model in this study, in which the parameters are linked to explainable social meanings. The parameter set $(N, k, p)$ is as follows.
\begin{itemize}
\item $N$: Number of nodes, namely the population in each residence or work network.
\item $k$: Each node joint with its $k$ nearest neighbors in a ring topology. $k$ describes the fundamental contacts of an individual in a network.
\item $p$: Probability of adding a new edge to each edge. The value range is $[0, 1]$.
\end{itemize}

After determining the parameter set, we generate Newman-Watts small-world network as follows.
First, a ring of $N$ nodes was generated. Each node connects to its $k$-nearest neighbors (or $k-1$ if $k$ is odd). 
Shortcuts are then added as new edges in the ring. The new edges are between the existing nodes chosen randomly by the parameter $p$. 
Do not remove any edges.

\subsection{Causal Inference of Social Infectivity}

Social infectivity indicates the infection effects of the propagation on a social level. Algorithm 2 illustrates the propagation mechanism, while any infectious node (E, A, or I) exists in the networks. Social infectivity is the interventions estimator used to conduct a causal analysis in the interconnected SEIRAH framework.

According to the SEIRAH mechanism, the hospitalized time series ($H_t$) has an evident monotonic relationship with a social infectivity value in a speciﬁc network. To deal with the computational procedures of the social infectivity ($\beta_t$), this study took $\beta_{t-1}$ (social infectivity value of one day before) as the predicted prior social infectivity to calculate the posterior social infectivity. Here, $t$ is the subscript of the time series (daily) variable. The predicted posterior social infectivity ($\beta_t$) was optimized until the preset accuracy was obtained in the model. The computational method is a modiﬁed binary search that applies the optimization processes that are described in Algorithm 1.

\begin{figure}[!t]
  \begin{algorithm}[H]
    \caption{Modified binary search for social infectivity time series ($\beta_t$)}
    \label{alg1}
    \begin{algorithmic}[1]
	\If {$E+A+I=0$}    
		\State $\widehat \beta^{t} \leftarrow 0$ 
      \Else
      		\State $\widehat \beta_{t} \leftarrow \beta_{t-1}, A \leftarrow 0, B \leftarrow 1$  
		\State $X \leftarrow \{\widehat \beta^{t}, A, B \}$  
		\While {$\left | A-B \right | >\epsilon$}   
		\State $D^{X} = \sum_{i=0}^{m} (\widehat{H}_{t+i}^{X} - H_{t+i})^2$ 
		\If {$D^{A} > D^{B}$}   
			\State $A \leftarrow \widehat \beta_{t}, D^{A} \leftarrow D^{\widehat \beta_{t}}$
	      \ElsIf {$D^{A} < D^{B}$}
			\State $B \leftarrow \widehat \beta_{t}, D^{B} \leftarrow D^{\widehat \beta_{t}}$
	      \EndIf 
	      \State $\widehat \beta_{t} \leftarrow \frac{1}{2}(A+B)$  
	      \EndWhile
      \EndIf
      \State \Return $\beta_{t}  \leftarrow \widehat \beta_{t}$
      
    \end{algorithmic}
  \end{algorithm}
\end{figure}

\section{Experiment and Results}

\subsection{Data Sources}

This study used an experimental dataset from the Tokyo metropolitan area. This dataset consisted of one central city (Tokyo) and three outskirts (Kanagawa, Saitama, and Chiba prefectures), as shown in Table \ref{tab_TokyoM}.

The Tokyo metropolitan area has a developed urban train system that serves around 83\% of the daily commuting people before the pandemic outbreak. The station peak-time indicator receives regular monitoring by the Japanese authority for social management. After the outbreak, Japan enhanced their monitoring for the number of train users during daily peak times and sampling from the main stations in the Tokyo and Osaka metropolitan areas.

\begin{figure*}[!t]
\centering
\includegraphics[width=1.0\textwidth]{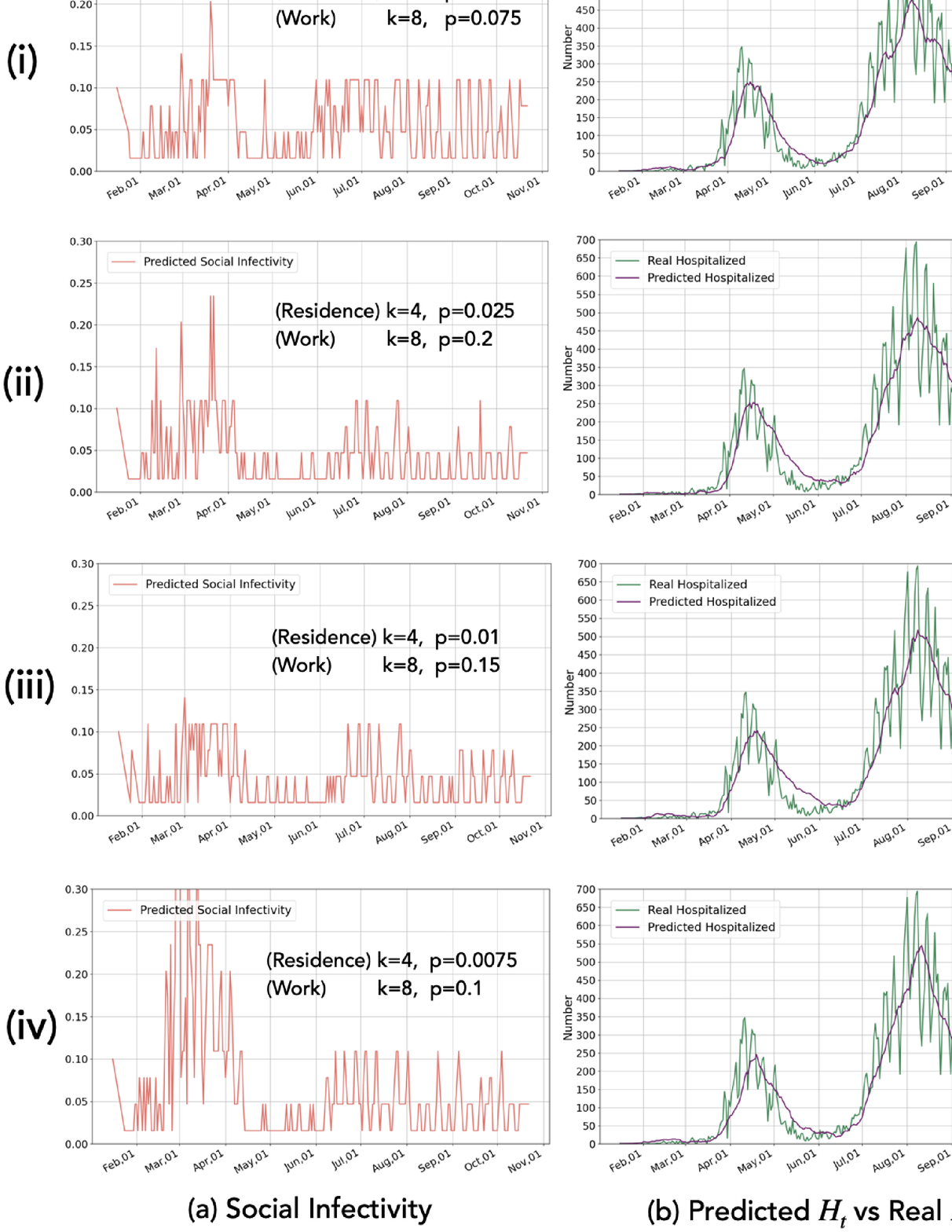}
\caption{Simulation results of the SEIRAH model in the interconnected networks with a variety of network patterns (i)(ii)(iii)(iv).}
\label{fig_Rst}
\end{figure*}

\begin{table}[!t] 
\renewcommand{\arraystretch}{1.3}
\caption{The population of the Tokyo metropolitan areas} 
\label{tab_TokyoM}
(unit: thousand people)
\centering
\begin{tabular}{l|rrrp{16mm}p{16mm}p{16mm}}\hline\hline  
\fontsize{8pt}{0pt}\selectfont City& \fontsize{8pt}{0pt}\selectfont Population & \fontsize{8pt}{0pt}\selectfont Daily Commuting* & \fontsize{8pt}{0pt}\selectfont \% of Population\\\hline
Tokyo &13,520 & 4,864 & 36.0\%\\
Kanagawa&9,200 &888 & 10.0\%\\
Saitama&7,340&780&10.6\%\\
Chiba&6,280&598&9.5\%\\\hline
Total&36,340&7,130&19.6\%\\\hline
\multicolumn{4}{l}{\fontsize{7pt}{0pt}\selectfont *People commute to the central city for daily work by public transportation.} 
\end{tabular}
\end{table}

\begin{table}[!t] 
\renewcommand{\arraystretch}{1.3}
\caption{Threshold values of status transition} 
\label{tab_threshold}
\centering
\begin{tabular}{l|rr}\hline 
\fontsize{8pt}{0pt}\selectfont Status Transition& \fontsize{8pt}{0pt}\selectfont Threshold& \fontsize{8pt}{0pt}\selectfont Calculated Value\\\hline
E $\to$ A &$p_{1}\sigma$ & 0.036 \\
E $\to$ I&$(1-p_{1}\sigma)$ & 0.164 \\
A $\to$ H&$p_2\lambda_{AH}$& 0.028\\
A $\to$ R&$(1-P_2)\gamma_{AR}$& 0.08\\
I $\to$ H&$\lambda_{IH}$& 0.950\\
H $\to$ R&$\gamma_{HR}$& 0.100 \\\hline
\multicolumn{3}{l}{\fontsize{7pt}{0pt}\selectfont *Parameter descriptions are in Fig.\ref{fig_SEIRAH}(a).}
\end{tabular}
\end{table}

\subsubsection{\textbf{Station Peak-time Indicator}}
An indicator of the Japanese authority sets a standard level, the average station user amount, which were monitored daily from February 24 to 28, 2020, just before the COVID-19 outbreak occurred in Tokyo. From March 1, this indicator is the daily user amount that is normalized by the mentioned standard level. In this study, this indicator calculated the daily commuting population.

\subsubsection{\textbf{Epidemiological Parameters}}

The SEIRAH model shown in Fig.\ref{fig_SEIRAH}(a) illustrates their relationships. They are summarized in Table \ref{tab_threshold} with the sources of public information from the Ministry of Health, Labour and Welfare (\url{www.mhlw.go.jp}, the Japan public health authorities), and the recently published studies \cite{Rt-2}\cite{A-p1}\cite{CDC_FDA}\cite{Microepidemics}. In this study, the parameters do not consider the new virus variations, which require more epidemiological evidence.

\subsection{Procedure and Results}

We generated four residence networks that have a 1/100 population scale to simulate Tokyo, Kanagawa, Saitama, and Chiba. One work network interconnected the four metropolitan areas, which was dynamically composed according to the daily commuting population. Here, the network parameter set $(N, k, p)$ is a particular pattern. The SEIRAH model was overlaid on these interconnected networks to explore the epidemic dynamics.

First, we used the hospitalized numbers to infer social infectivity in time series, namely $\beta_t$ in the generated networks.

Second, we tested the various network parameter patterns to search for the appropriate topology by the parameter p to ﬁt the experimental data well in the simulation.

Finally, we obtained the time series of the status transition by using the above procedures, which is the target information that concerns the propagation features to explore the epidemic dynamics.

\subsubsection{\textbf{Social infectivity ($\beta_t$)}}
In the early days of the outbreak, the social infectivity was relatively high. This is demonstrated in Fig.\ref{fig_Rst}(a), which shows the lack of public awareness in the COVID-19 pandemic. While the situation worsened, the government announced a state of emergency. Most of the people started to pay serious attention to social distancing and their outside movements. At this time, the severity of the social infections decreased. The simulation results are as follows.

\begin{itemize}
\item The recent social infectivity has shown a $\beta_t < 0.1$. $\beta_t$ decreases to approximately 30\% from the early outbreak period. This trend indicates a mindful awareness of the public health in the Tokyo metropolitan area.
\item We control the propagation under the condition of $\beta_t \to 0$, which is also the target of public health policy.
\item The social infectivity changes signiﬁcantly before and after the state of emergency in Japan. The behavior restriction causes the social infectivity to decrease.
\item Non-pharmaceutical interventions (NPIs) affect the propagation; however, it is difficult to isolate the infections completely. This fact tends to show that only pharmaceutical interventions can terminate the pandemic on a social level.
\end{itemize}

\subsubsection{\textbf{Network parameters ($p^{(r)}$, $p^{(w)}$)}}

\begin{figure*}[!t]
\centering
\includegraphics[width=1.0\textwidth]{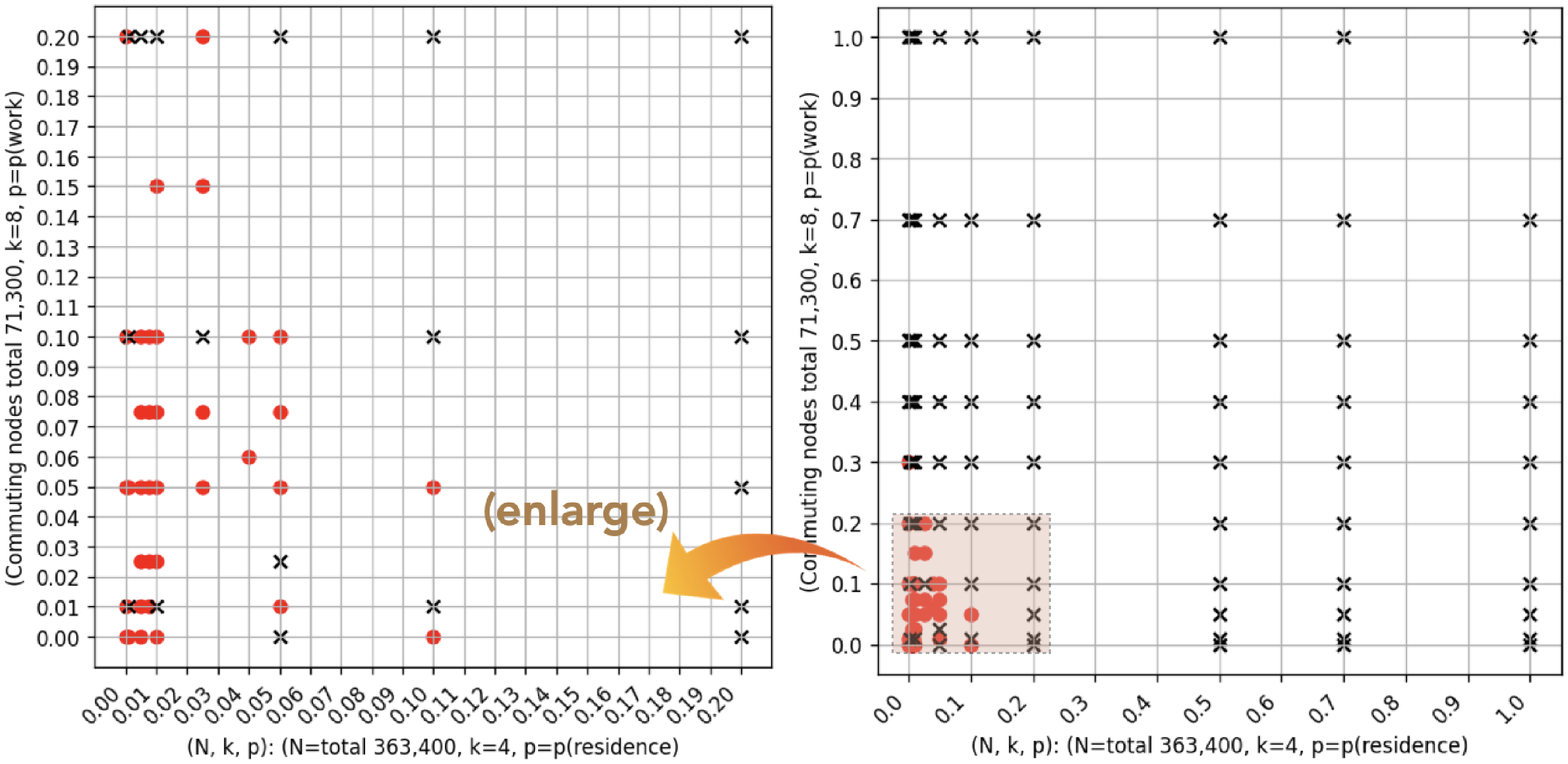}
\caption{
Parameter $p$ patterns of the residence networks and work network, namely the pairs of  $p^{(r)}$ and  $p^{(w)}$, for the 363,400 node-size interconnected network simulation. The `\textbf{$\times$}' indicates the tested inappropriate network topology, and the `red \textcolor{red}{$\bullet$}' indicates the appropriate ones.}
\label{fig_p_pattern}
\end{figure*}

A small-world network has the parameter set of $(N, k, p)$, as previously mentioned in Section III.B. The small-world model initializes a flexible topology to simulate the real world by adjusting the network topology. Fig.\ref{fig_Rst} shows the simulation results for the different network parameter sets.

Different network topologies may cause different epidemic dynamics \cite{HiddenGeometry}\cite{NWK-Science}. This study explores the different network topologies by adjusting the parameter $p$. The causality of $p$ and $H_t$ is challenging to explore. Thus, we tested the various patterns to search for a suitable range or value of $p$ by Algorithm 1. A suitable $p$ outputs the predicted $H_t$, which matches the real $H_t$ well.

Our trial contained a total of 363,400 node-scale interconnected networks. Here, we abbreviated the parameter $p$ of the residence network as $p^{(r)}$, and the work network is abbreviated as $p^{(w)}$. The trial steps were as follows.

(1) Set patterns of p(r) and p(w) in the range of $[0, 1]$.

(2) Input the $p$ pattern and generate the networks for the simulation.

(3) Evaluate the output of the social infectivity $\beta_t$ and predicted $H_t$ time series.

 We trailed a variety of $p$ patterns as shown in Fig.\ref{fig_p_pattern}, with some ﬁndings as follows.
\begin{itemize}
\item The appropriate network patterns were estimated in the range of $p^{(r)} \in [0, 0.1]$ and  $p^{(w)} \in [0, 0.2]$.
\item The signiﬁcant ranges are $p^{(r)} \in [0, 0.05]$ and  $p^{(w)} \in [0, 0.1]$. 
\item In most cases, $p^{(r)} < p^ {(w)}$. This indicates that out-family contacts are always more frequent than within-family contacts in the social networks.
\end{itemize}

\section{Discussion}

\subsection{\textbf{Interpretability of the Observed Data}}

Interpretability is an essential, yet an often-confusing issue, when applying the analysis models to the real world because of the known and unknown bias or causality. The non-pharmaceutical intervention (NPI) analysis is a popular method that has been performed in recent studies. We noticed a topic that has emerged in a few papers published in Nature \cite{Flaxman-1}\cite{Soltesz}\cite{Flaxman-2}  in terms of estimating the effects of the NPIs on COVID-19 in Europe. 
 
In this emerging topic, by focusing on the selected five NPIs (social distancing encouraged, self-isolation, school closures, public events banned, and complete lockdown), Flaxman et al. \cite{Flaxman-1} pointed out that the lockdown had the most identifiable effect on the propagation reduction. Soltesz et al. \cite{Soltesz} examined this result with the same model and indicated that the invention of the banned public events has a more identifiable effect than the complete lockdowns. After a further analysis, Flaxman et al. \cite{Flaxman-2} noted that the different model pooling methods resulted in different results. Flaxman et al. utilized a partial pool of 11 European countries. However, Soltesz et al. used a full pool for the individual Swedish data.
 
From these studies, we know that the interventions' effect estimation may cause a bias and different interpretations while neglecting the regional scale and causal inference. First, we chose a metropolitan scale, which is smaller. As a result, the metropolitan scale is better for interventional bias control than the country and multi-country scales. We can then represent the social infectivity, which is an average estimator of the COVID-19 transmission that covers the effects of the defined and undefined NPIs. The social infectivity shown in Fig.\ref{fig_Rst}(a) is an outcome of the mechanism from our epidemic model, which applies to causal interpretation.

In our study, the epidemic model outputs the simulated hospitalized number series ($H_t$) in Fig.\ref{fig_Rst}(b) on a certain level of social infectivity. The simulated $H_t$ can be compared with the observed real $H_t$ to control the simulation bias.

Many studies use the hospitalized number $H_t$ series as a data source by performing a parametric or non-parametric analysis. Ren et al. \cite{SSA-GF} presented a non-parametric analysis of the singular spectral analysis (SSA). The SSA decomposes the hospitalized number series into different components without prior assumptions. As a model-free technique, the SSA avoids the empirical parameters, and it has a good computing efficiency.

Unlike SSA, our study adopted a parametric analysis because of its emphasis on data interpretability. Our model outputs nearly the same results in Fig.\ref{fig_Rst}(b) as demonstrated in the SSA method by Ren et al. Moreover, with the propagation mechanism in the epidemic model, our model builds an interpretable path from the infectivity, as shown in Fig.\ref{fig_Rst}(a), to the transition of the infectious classes, as shown in Fig.\ref{fig_Rst}(c). Our method improves the data interpretability, which is an essential advantage for propagation understanding and decision making.

Meanwhile, as a parametric analysis, our study is limited by the empirical parameters. Recently, viral variations have been confirmed in many countries. In terms of future work, we will need to reinforce our model by updating the necessary parameters with more epidemiological evidence.

\subsection{\textbf{Propagation Feature Findings of the Infectious Classes}}

\subsubsection{\textbf{The peak of the exposed (E) individuals emerges two weeks ahead of `H'.}}

In Fig.\ref{fig_Rst}(c), we can find two peaks with `E' numbers at the beginning of April and August. Then, the `H' peaks follow after two weeks. According to the World Health Organization (WHO), the incubation period of COVID-19 ranges from 1 to 14 days with a median of 5 days. The situation in Tokyo confirms this. This propagation feature finding emphasizes the importance of checking in the target regions or persons. Except when performing polymerase chain reaction (PCR) testing in a hospital, a simple home checking kit might be worthier of being promoted. To confirm the rising exposed ratio, more buffer days were required to enhance the local medical measurements.

\subsubsection{\textbf{Asymptomatic (A) has a subsequent descent after `H'.}}
This must be considered when controlling COVID-19. Another propagation feature finding is that the trend transition of `A' follows `H' with a signiﬁcant subsequent descent. The `H' descent quickly causes the overhasty judgment of the `getting better' situation. However, this fact is not positive because of the unobservable `A' of the vast majority. In Japan, once the number of hospitalizations decreased in May, the authority lifted the state of emergency at the end of May and soon promoted the `Go To Eat' and `Go To Travel' policy for economic revitalization. This overhasty policy accelerated human mobility and triggered a second outbreak in July. Meanwhile, we have seen similar overhasty policies in other countries.

The transition of the `A' population also emphasizes the importance of checking. In the SEIRAH mechanism, by having more checking, the more asymptomatic infected individuals will be inspected and accelerated to the `H' class. As a result, we can protect more susceptible individuals from social networks.

\subsubsection{\textbf{The symptomatic infected (I) has a faster trend than `A'.}}
In the SEIRAH model, the `E' individuals transit to `I' stochastically as a causal effect. In Fig.\ref{fig_Rst}(c), we can see that the trend of `I' is similar to `A'; however, it decreases relatively fast. The Tokyo metropolitan areas have developed medical systems with hierarchical hospitals and community clinics. This system accelerates the transition of `I' to `H', while the symptoms appear. Meanwhile, we can use the ratio of `I' in the total `H' from the hospital statistics to estimate the `A' population. Controlling `A' might be more important than `I' with regard to public health authority tasks.

\section{Conclusion and Future work}

\subsection{Conclusion} 

First, this study extends the classic epidemic model with the new features of COVID-19. This study reinforced the existing stochastic processing analysis of the propagation by combining the causal interpretation and overlaying the model on the networks. This method improved the practical level of the epidemic model that has conditions that simulate the actual conditions in a society.

Second, recent studies have mainly focused on predicting the trend of the daily hospitalized numbers and they have overlooked the mechanism of the infection causes. This study contributes to the core mechanism of the causal inference, which focuses on the average effect of the non-pharmaceutical interventions. This is named as the social infectivity in the model, which illustrates an interpretable path between the observed effect and the unobservable causes of the propagation. 

Third, we highlighted the interconnected networks to understand the epidemic dynamics while considering human mobility effects, which is the first factor of propagation in modern society. This study explored the epidemic dynamics spatiotemporally by using the Tokyo metropolitan dataset. This has interpretable outcomes of the unobservable social factors that contains new feature findings. This can help us obtain a better understanding of COVID-19 and provide us with valuable local references for propagation control.

\subsection{Future Work}

The stochastic processes method can provide more reliable results when the amount of network nodes is closer to the target region's population. Theoretically, the computational complexity is between $O(N)$ and $O(NlogN)$ by the current network traversal. A larger network size results in a much longer computational time. Therefore, the algorithms in this study are limited to a vast network, such as millions of nodes or more. We expect that the graph neural network (GNN) is an advanced option to ﬁt our networked structure with a better data analyzing performance.

Some recent studies have challenged the development of more computational methods and algorithms that concerns the causal analysis in the GNN or networked structures that are similar. This direction is expected to explore more features from observable information and reinforce the interpretability of the data analysis.

Moreover, these models are expected to implement more counterfactual experiments with signiﬁcant social implications. COVID-19 is a new pandemic to mankind with many unknown characteristics. Meanwhile, we are limited to evaluating new control measurements by performing social testing. From this, 
counterfactual experiments can explore more epidemic dynamics for decision making, which is a primary task in our future work.



\begin{thebibliography}{99}

\bibitem{Newman2006}
Newman, Mark \& Barabasi, Albert-Laszlo \& Watts, Duncan. (2006). The Structure and Dynamics of Networks. Princeton University Press. ISBN:978-0-691-11357-9

\bibitem{Gama}
Telo da Gama,M., Nunes,A. Epidemics in small world networks. Eur. Phys. J. B 50, 205–208(2006).

\bibitem{WS1998}
Watts,D., Strogatz,S. Collective dynamics of `small-world' networks. Nature 393,440- 442 (1998).

\bibitem{Barrat}
Barrat, A., Weigt, M. On the properties of small world network models. Eur. Phys. J. B 13, 547–560 (2000). 

\bibitem{HiddenGeometry}
Brokmann D et al, The Hidden Geometry of Complex, Network-Driven Contagion Phenomena. 
Science 14 Feb 2014: Vol. 343, Issue 6172, pp. 730

\bibitem{PLos}
Liu M et al. (2015), Epidemics in Interconnected Small-World Networks. PLoS ONE 10(3):e012070.

\bibitem{NWK-Science}
Albert-Laszlo Barabasi, Network Science, July 21, 2016. Cambridge University Press. ISBN-13: 978-1107076266

\bibitem{Transit}
X.Qian, L.Sun and S.V.Ukkusuri, Scaling of contact networks for epidemic spreading in urban transit systems, Feb.,10, 2020. 

\bibitem{Mobility}
Chang, S. et al. Mobility network models of COVID-19 explain inequities and inform reopening. Nature (2020).

\bibitem{SW-SARS}
Small,Michael and Tse,Chi. (2005). Small World and Scale free Model of Transmission of SARS. I. J. Bifurcation and Chaos. 15. 1745-1755. 

\bibitem{Kucharski}
A. J. Kucharski et al., Early dynamics of transmission and control of COVID-19: A mathematical modeling study, Lancet Infect. Dis., vol. 20, no. 5, pp. 553–558, May 2020

\bibitem{Humphries}
Humphries MD, Gurney K (2008) Network `Small-World-Ness': A Quantitative Method for Determining Canonical Network Equivalence. PLoS ONE 3(4): e0002051.

\bibitem{Causal_1}
David Kaplan, Causal Inference for Observational Studies, The Journal of Infectious Diseases, Volume 219, Issue 1, 1 January 2019, Pages 1–2

\bibitem{SD-1}
Reluga, Timothy. (2010). Game Theory of Social Distancing in Response to an Epidemic. PLoS computational biology.

\bibitem{Threshold-1}
Wang Huijuan, et al. (2013, 2018). Effect of the interconnected network structure on the epidemic threshold. Physical review. E, Statistical, nonlinear, and soft matter physics. 88. 022801

\bibitem{SSA-GF}
Ren, Jinchang et al. A Novel Intelligent Computational Approach to Model Epidemiological Trends and Assess the Impact of Non-Pharmacological Interventions for COVID-19. IEEE J Biomed Health Inform ; 24(12): 3551-3563, 2020 12.

\bibitem{EpiWave}
Hoen AG et al. Epidemic Wave Dynamics Attributable to Urban Community Structure: A Theoretical Characterization of Disease Transmission in a Large Network. J Med Internet Res 2015;17(7):e169

\bibitem{Bianconi}
Bianconi, Ginestra. (2016). Epidemic spreading and bond percolation in multilayer networks. Journal of Statistical Mechanics: Theory and Experiment. 2017. 

\bibitem{Wei}
Wei, Xiang et al.(2017). Cooperative Epidemic Spreading on a Two-Layered Interconnected Network. SIAM Journal on Applied Dynamical Systems. 

\bibitem{Lee}
C.Lee, S.Tenneti, D.Y.Eun., Transient Dynamics of Epidemic Spreading and Its Mitigation on Large Networks. Proc. ACM MobiHoc 2019

\bibitem{Wu}
Wu, Yulin and Cai, Wentong et al. Efficient Parallel Simulation over Large-Scale Social Contact Networks (April 2019), ACM Transactions on Modeling and Computer Simulation. 

\bibitem{Rt-1}
Anne Cori et al. A New Framework and Software to Estimate Time-Varying Reproduction Numbers During Epidemic, American Journal of Epidemiology, Volume 178, Issue 9, 1 November 2013, Pages 1505–1512 

\bibitem{Rt-2}
Qifang Bi et al. Epidemiology and transmission of COVID-19 in 391 cases and 1286 of their close contacts in Shenzhen, China: a retrospective cohort study. The LANCET, April 27, 2020 

\bibitem{Eletreby}
Eletreby Rashad, Zhuang Yong et al. (2020). The effects of evolutionary adaptations on spreading processes in complex networks. Proceedings of the National Academy of Sciences.

\bibitem{SocialPhy}
Bhattacharya, Kunal. Social physics: uncovering human behavior from communication. Advances in physics: X. 2019, vol. 4, no. 1. 

\bibitem{A-p1}
Mizumoto Kenji et al. Estimating the asymptomatic proportion of coronavirus disease 2019 (COVID-19) cases on board the Diamond Princess cruise ship, Yokohama, Japan, 2020. Euro Surveill. 2020

\bibitem{Review_M_Agent}
Herrera, Manuel. A Review on Control and Optimization of Multi-Agent Systems and Complex Networks for Systems Engineering. Preprints. 2020.
  
\bibitem{China}
Qin, Lei. A Network Analysis of COVID-19 Epidemic in Mainland China by K-Core Decomposition. JMIR public health and surveillance. 2020.
  
\bibitem{CDC_FDA}
A Network-Based Stochastic Epidemic Simulator: Controlling COVID-19 with Region-Specific Policies. Medical Letter on the CDC \& FDA. 2020.

\bibitem{Italy}
Della Rossa et al. A network model of Italy shows that intermittent regional strategies can alleviate the COVID-19 epidemic. Nature communications. 2020, vol. 11, no.1, p.5106–5106. 
  
\bibitem{Alexander}
Karaivanov, Alexander. A social network model of COVID-19.(Research Article)(Report). PLoS ONE. 2020, vol.15, no.10, p.e0240878.

\bibitem{Modelling}
Small, Michael. Modelling Strong Control Measures for Epidemic Propagation With Networks-A COVID-19 Case Study. IEEE access. 2020, vol. 8, p. 109719–109731.
  
\bibitem{Microepidemics}
Mylona, Evangelia K. Real-Time Spatiotemporal Analysis of Microepidemics of Influenza and COVID-19 Based on Hospital Network Data: Colocalization of Neighborhood-Level Hotspots. American journal of public health. 2020, p. e1–e8.

\bibitem{Social_Interactions}
Herrmann, Helena A. Why covid-19 models should incorporate the network of social interactions. Physical Biology. 2020, vol.17, no.6, p.9.

\bibitem{SEIRD}
Fenglin Liu et al. Predicting and analyzing the COVID-19 epidemic in China: Based on SEIRD, LSTM and GWR models. PloS one. 2020, vol. 15, no. 8, p. e0238280. 

\bibitem{Metro_1}
Kumar, Nishant. Activity-based contact network scaling and epidemic propagation in metropolitan areas. 2020.

\bibitem{Social_Networks}
Piccardi, C. Social networks and the spread of epidemics. Lett Mat Int 1, 119–126 (2013).

\bibitem{Flaxman-1}
Flaxman, S., Mishra, S., Gandy, A. et al. Estimating the effects of non-pharmaceutical interventions on COVID-19 in Europe. Nature 584, 257–261 (2020). 

\bibitem{Soltesz}
Soltesz, K., Gustafsson, F., Timpka, T. et al. The effect of interventions on COVID-19. Nature 588, E26–E28 (2020). 

\bibitem{Flaxman-2}
Flaxman, S., Mishra, S., Scott, J. et al. Reply to: The effect of interventions on COVID-19. Nature 588, E29–E32 (2020). 

\end{thebibliography}
\end{document}